\documentclass[11pt]{article}

\usepackage[T1]{fontenc}
\usepackage[utf8]{inputenc}
\usepackage[margin=1in]{geometry}
\usepackage{graphicx}
\usepackage{amsmath}
\usepackage[bottom]{footmisc}
\usepackage{ragged2e}
\usepackage{microtype}
\usepackage{parskip}
\usepackage{hyperref}
\usepackage{booktabs}
\usepackage{float}
\usepackage{fancyvrb}

\title{Towards Operational Conversational Intelligence:\\A Speech Intelligence Framework}
\author{\textbf{Chaitanya Vishnoi}\thanks{Work done during internship at EXL}\\Indian Institute of Technology Kanpur\\\texttt{chaitanyav23@iitk.ac.in}\and \textbf{Shudhant Khurana}\\EXL\\\texttt{Shudhant.Khurana@exlservice.com}\and \textbf{Abhilash Timmapur}\\EXL\\\texttt{Abhilash.Timmapur@exlservice.com}\and \textbf{ Somya Rai}\\EXL\\\texttt{Somya.Rai@exlservice.com}\and \textbf{Saket Mohanty}\\EXL\\\texttt{Saket.Mohanty@exlservice.com}}
\date{July 2026}

\begin{document}

\justifying

\maketitle


\begin{abstract}

\noindent Body-worn camera (BWC) audio presents unique challenges---high ambient noise, variable recording conditions, and multiple overlapping speakers---that make automated transcription and speaker labeling very difficult. We propose a \textbf{dual-path conversational intelligence framework} that preprocesses raw BWC audio, splits the task into a diarization branch and an ASR branch, and fuses their outputs. The diarization branch uses a denoising front-end (DeepFilterNet) and voice activity detection (VAD) to identify speaker turns, followed by NVIDIA's Multi-Scale Speaker Diarization Decoder (MSDD) with TitaNet embeddings. The transcription branch uses loudness normalization and WhisperX (Large-v3) with forced alignment and probability-guided speech segmentation. Finally, we perform word-level speaker attribution by assigning each recognized word to the speaker segment with the maximum temporal overlap. We evaluate the proposed framework on a curated body-worn camera dataset
constructed from publicly available U.S. and U.K. police body-worn camera
recordings. Experimental results demonstrate that task-specific acoustic conditioning and probability-guided speech segmentation improve speaker diarization, transcription, and word-level speaker attribution under challenging body-worn camera recording conditions. The proposed modular architecture further provides an extensible foundation for future speaker-aware conversational intelligence systems.

\end{abstract}

\vspace{0.5em}


\newpage
\section{Introduction}

Body-worn cameras (BWCs) are increasingly deployed by law enforcement agencies, generating large volumes of multimodal evidence from police--civilian interactions \cite{BJS2016,NIJVideoAnalysis}. Extracting meaningful information from these recordings through automatic transcription and speaker identification can significantly improve evidence review, officer training, incident analysis, and investigative workflows \cite{NIJVideoAnalysis,McCluskey2023}.  

\noindent However, BWC audio is considerably more challenging to process than conventional speech corpora. Recordings frequently contain environmental noise, overlapping speech, reverberation, microphone movement, sirens, traffic noise, radio communication, and rapidly varying recording conditions. These factors substantially degrade the performance of both automatic speech recognition (ASR) and speaker diarization systems by reducing transcription accuracy and increasing speaker confusion \cite{Khazaleh2024,Godin2015,Charlet2013}.

\noindent Recent advances have produced highly capable models for the individual components of conversational speech processing. Whisper provides robust large-scale speech recognition, while WhisperX extends Whisper through voice activity detection and forced alignment to produce accurate word-level timestamps. Similarly, the Pyannote framework provides state-of-the-art neural voice activity detection and speaker diarization components, and NVIDIA's Multi-Scale Speaker Diarization Decoder (MSDD) achieves competitive speaker clustering through multi-scale speaker representations \cite{Whisper,WhisperX,Pyannote,MSDD}. Nevertheless, these models are typically optimized independently and are not designed as a unified processing pipeline specifically for noisy body-worn camera recordings.

\noindent To address this limitation, we propose an end-to-end \textbf{Conversational Intelligence} framework for BWC audio. Our approach follows a dual-path architecture in which the diarization and ASR branches employ different preprocessing strategies that are individually optimized for their respective tasks. The diarization branch performs DeepFilterNet-based speech enhancement followed by the proposed probability-guided Pyannote voice activity detection pipeline and multi-scale speaker diarization, whereas the ASR branch performs loudness normalization prior to WhisperX transcription and forced alignment. Finally, a deterministic maximum-overlap fusion algorithm combines the outputs of both branches by assigning each recognized word to the speaker segment exhibiting the greatest temporal overlap.

\noindent The proposed framework adopts a modular and reproducible architecture in which every processing stage produces auditable intermediate artifacts. This enables independent evaluation of voice activity detection, speaker diarization, automatic speech recognition, and speaker attribution while facilitating future replacement or improvement of individual components without modifying the remainder of the pipeline.

\noindent The main contributions of this work are summarized as follows:

\begin{itemize}
    \item A \textbf{modular end-to-end framework for speaker-aware conversational intelligence} from body-worn camera recordings, integrating voice activity detection, speaker diarization, automatic speech recognition, and speaker attribution.
    
    \item A \textbf{dual-path acoustic conditioning strategy} that independently optimizes the preprocessing requirements of speaker diarization and automatic speech recognition while maintaining a shared temporal grounding.
    
    \item A \textbf{probability-guided voice activity detection post-processing algorithm} that performs recursive speech segmentation based on speech posterior probabilities to generate WhisperX-compatible speech segments.
    
    \item A \textbf{deterministic word-level speaker-attribution} algorithm that integrates WhisperX forced alignment with speaker diarization using maximum temporal overlap.
    
    \item A comprehensive experimental evaluation on a manually annotated body-worn camera dataset, including quantitative analysis of voice activity detection, speaker diarization, automatic speech recognition, and speaker-attribution performance together with architectural ablation studies (Appendix~\ref{appendix:ablation}).
\end{itemize}

\section{Research Gap and Motivation}

Unlike controlled speech corpora, body-worn camera (BWC) recordings contain noisy far-field speech, overlapping conversations, highly variable acoustic environments, reverberation, microphone movement, and diverse speaking styles, making both automatic speech recognition and speaker diarization considerably more challenging \cite{Khazaleh2024,Godin2015,Charlet2013,McCluskey2023}.

\noindent To the best of our knowledge, there is currently no large publicly available benchmark specifically designed for end-to-end conversational intelligence on police body-worn camera audio. Although several widely used speech corpora, including \textit{LibriSpeech, AMI, CALLHOME, VoxConverse,} and \textit{AISHELL}, have significantly advanced research in speech recognition and speaker diarization, they do not adequately represent the acoustic conditions encountered in operational BWC deployments \cite{LibriSpeech,AMI,CALLHOME,VoxConverse,AISHELL}.

\noindent This lack of publicly available benchmark datasets motivated the creation of our own evaluation dataset, comprising publicly available BWC-style recordings together with manually generated transcripts and speaker annotations. The dataset was designed to capture conversational characteristics and acoustic conditions commonly encountered in body-worn camera recordings while enabling stage-wise evaluation of the proposed framework.

\noindent Another practical challenge arises from the conflicting acoustic requirements of automatic speech recognition and speaker diarization. Aggressive speech enhancement can improve downstream speaker diarization by suppressing background interference, yet the same enhancement may distort acoustic cues that are important for robust speech recognition, thereby increasing transcription errors. In contrast, preprocessing strategies that preserve lexical information for ASR do not necessarily produce speaker representations that are optimal for diarization.

\noindent These observations motivate the proposed dual-path framework, where each branch employs task-specific preprocessing while preserving a common temporal representation that enables accurate fusion during the orchestration stage.

\noindent No existing open-source framework provides an end-to-end pipeline that jointly combines the components investigated in this work while specifically targeting conversational intelligence for body-worn camera recordings. We are unaware of an open-source framework that integrates task-specific acoustic conditioning, neural voice activity detection, speaker diarization, WhisperX transcription and forced alignment, and temporal speaker-attribution fusion into a unified pipeline specifically designed for body-worn camera audio.

\noindent Several conversational and emotion-recognition datasets, including \textit{IEMOCAP, MSP-Podcast,} and \textit{MELD}, provide valuable multi-speaker speech recordings with rich annotations for emotion and dialogue analysis. However, these datasets were not designed for the acoustic conditions encountered in body-worn camera recordings, lacking characteristics such as severe environmental noise, radio communication, rapid speaker movement, and highly variable recording conditions. Consequently, they do not adequately represent the acoustic conditions encountered in operational BWC recordings and are therefore not ideal benchmarks for evaluating end-to-end conversational intelligence pipelines in this domain. \cite{IEMOCAP,MSPPodcast,MELD}.

\section{Data}

\subsection{Data Collection}

To evaluate the proposed framework, we assembled a small body-worn camera (BWC)-like dataset from publicly available English-language recordings. Since, to the best of our knowledge, no standardized public benchmark currently exists for end-to-end conversational intelligence on police body-worn camera audio, videos were collected from publicly accessible YouTube channels that publish body-worn camera footage, including official law enforcement agencies, government organizations, major news organizations, and publicly released police recordings \cite{NIJVideoAnalysis,BJS2016,CBPBWC}. The resulting dataset comprises \textbf{eight recordings} covering diverse conversational scenarios and acoustic environments representative of real-world BWC deployments. A summary of the dataset characteristics and provenance is provided in Tables~\ref{tab:dataset_summary} and~\ref{tab:dataset_provenance}.

\noindent The collected recordings were used exclusively for research and evaluation purposes, and all annotations were generated manually by the authors. Each recording was manually annotated to produce manually verified speaker-aware transcripts containing timestamped speaker turns. These annotations served as the ground truth for both automatic speech recognition (ASR) and speaker diarization evaluation.

\noindent An example annotation is shown below.

\begin{center}
\begin{BVerbatim}
Speaker_0 [00:00:00 - 00:00:02] I think they're looking for us.
Speaker_1 [00:00:03 - 00:00:03] We are, yes.
\end{BVerbatim}
\end{center}

\noindent To facilitate reproducible evaluation, the annotations were converted into the standard formats required by the downstream evaluation tools, including RTTM files for speaker diarization and timestamped word-level reference files for automatic speech recognition evaluation.

\subsection{Dataset Characteristics}

The evaluation dataset consists of eight publicly available English-language conversational recordings selected to represent diverse real-world acoustic conditions encountered in body-worn camera (BWC) applications as well as related conversational environments. The selected recordings comprise police body-worn camera footage, parliamentary debates, and broadcast interviews, thereby providing substantial variability in recording conditions, speaker interactions, and background noise.

\noindent In total, the evaluation corpus contains eight recordings with a combined duration of approximately 31 minutes and 31 seconds. Across all recordings, a total of 41 speaker instances were manually annotated, corresponding to an average of 5.13 speakers per recording. All recordings were converted to 16 kHz mono audio before processing and manually annotated to generate speaker-aware timestamped transcripts that served as the reference for both automatic speech recognition and speaker diarization evaluation.

\noindent The recordings intentionally span a wide range of acoustic environments including outdoor police interactions, aircraft cabins, television studios, parliamentary sessions, and roadside encounters. Common background disturbances include wind noise, public chatter, traffic, emergency sirens, aircraft engine noise, camera clicks, handcuff sounds, and overlapping speech. This diversity enables evaluation under challenging real-world conditions rather than laboratory-quality speech recordings.

\begin{table}[H]
\centering
\caption{Summary of the evaluation dataset.}
\label{tab:dataset_summary}
\begin{tabular}{lc}
\toprule
Statistic & Value\\
\midrule
Number of recordings & 8\\
Total duration & 31 min 31 s\\
Total annotated speaker instances & 41\\
Average speakers per recording & 5.13\\
Sampling rate & 16 kHz (mono)\\
Language & English (US and UK)\\
Ground truth & Manual speaker-aware transcripts\\
\bottomrule
\end{tabular}
\end{table}

\begin{table}[H]
\centering
\caption{Dataset provenance and recording characteristics.}
\label{tab:dataset_provenance}
\begin{tabular}{p{3.8cm}c c c p{2cm} p{4.2cm}}
\toprule
Recording & Duration & Speakers & Environment & Device & Dominant Acoustic Conditions\\
\midrule
Bodycam footage show arrests of Sara Sharif's alleged killers & 1m40s & 5 & Hybrid & BWC & Public chatter, overlapping speech\\
\hline
Dennis Skinner kicked out of Commons & 2m56s & 6 & Indoor & Professional microphone & Public chatter, overlapping speakers\\
\hline
Karen Has A Problem With Disabled Parking & 3m26s & 5 & Outdoor & Phone camera & Wind, overlapping speech\\
\hline
Piers Confronts Tommy Robinson & 3m25s & 3 & Indoor & Studio microphone & Broadcast discussion, overlapping speech\\
\hline
Police bodycam shows arrest of Southwest Airlines pilot & 2m42s & 4 & Aircraft cabin & BWC & Aircraft engine, wind, handcuff noise\\
\hline
Rich Karen Messes Around \& Finds Out & 10m52s & 9 & Outdoor & BWC & Sirens, wind, dog barking, glass breaking, camera clicks\\
\hline
That was a bit of a cheap shot weren't it mate & 2m55s & 4 & Outdoor & BWC & Wind, overlapping speech\\
\hline
UK Parliament Debate & 3m35s & 5 & Indoor & Professional microphone & Public chatter, parliamentary debate\\
\bottomrule
\end{tabular}
\end{table}

\subsection{Pre-processing}

Prior to model inference, several audio conditioning strategies were investigated to determine their suitability for speaker diarization and automatic speech recognition. Conventional preprocessing techniques, including RMS normalization and peak normalization, were found to provide only limited improvements under challenging body-worn camera recording conditions. Consequently, the proposed framework employs task-specific acoustic conditioning tailored to the requirements of each processing branch.

\begin{figure}[h]
    \centering
    \includegraphics[width=0.49\textwidth]{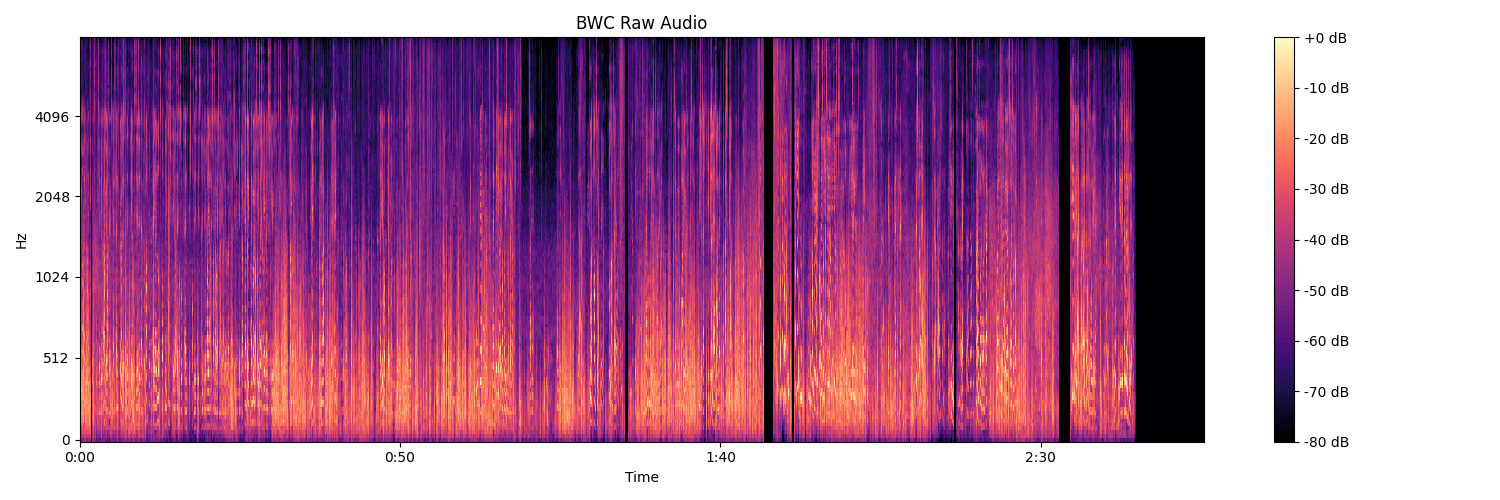}
    \includegraphics[width=0.49\textwidth]{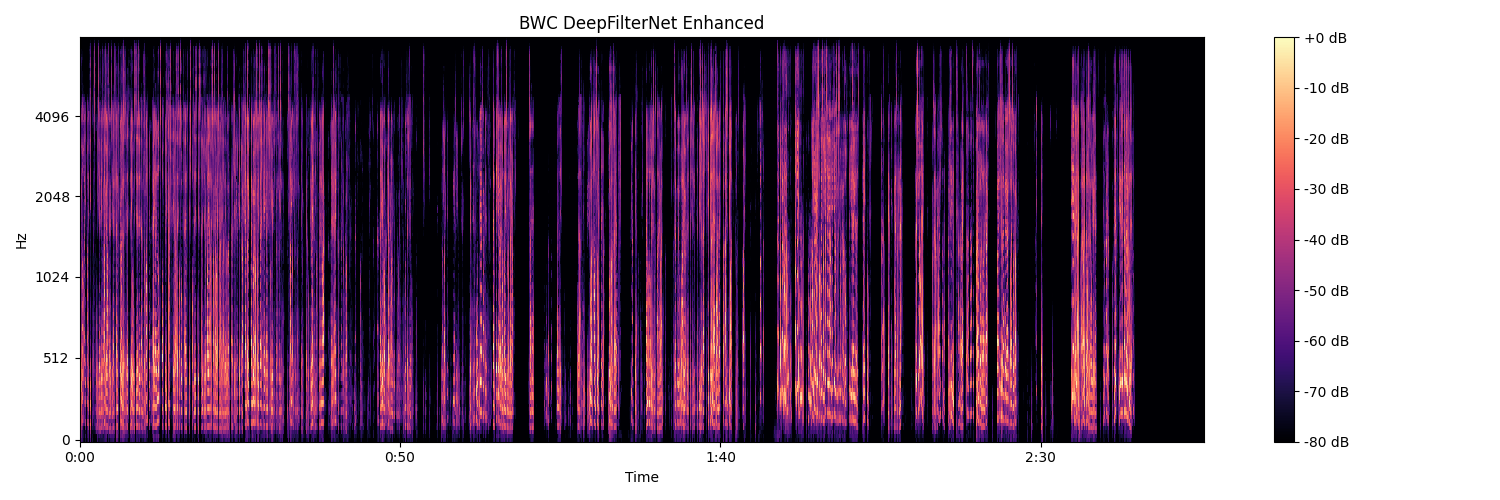}
    \caption{Mel-spectrogram comparison of a raw body-worn camera recording (left) and the corresponding DeepFilterNet-enhanced recording (right). DeepFilterNet suppresses low-frequency environmental noise while largely preserving the dominant speech components used for speaker diarization.}
    \label{fig:dfn_spec}
\end{figure}

\noindent For the diarization branch, DeepFilterNet was selected as the primary speech enhancement method owing to its ability to suppress environmental noise while preserving speech intelligibility \cite{DeepFilterNet}. Figure~\ref{fig:dfn_spec} compares the mel-spectrogram of a raw body-worn camera recording with its DeepFilterNet-enhanced counterpart. The enhanced spectrogram exhibits a noticeable reduction in low-frequency background noise and non-speech energy while preserving the dominant speech harmonics, providing a cleaner input signal for the subsequent speaker diarization pipeline.

\noindent The conflicting behaviour observed during the preliminary experiments revealed
what we refer to throughout this work as the \textbf{Enhancement Trap}. Although neural speech enhancement produces audio that is perceptually cleaner for human listeners, it simultaneously alters spectral and temporal characteristics of the speech signal, including harmonic structure, transient information, and high-frequency phonetic cues. Modern ASR models, including Whisper, are trained on large-scale speech
corpora containing substantial acoustic variability. Consequently, aggressive neural speech enhancement may alter acoustic characteristics that are informative for the ASR model, potentially reducing transcription accuracy despite improving perceptual speech quality. This behaviour was consistently observed during our experiments. In contrast, speaker diarization generally benefits from the suppression of
background noise because cleaner speech can provide more reliable inputs for
speaker representation learning~\cite{TitaNet,XVector} and speech activity
detection. The observed trade-off between speaker diarization and automatic speech recognition forms the basis of the proposed dual-path architecture, in which speech enhancement is applied exclusively within the diarization branch, while the transcription branch preserves the original acoustic characteristics through loudness normalization alone.
Representative acoustic conditioning experiments supporting this design choice
are presented in Appendix~\ref{appendix:ablation}.

\noindent Accordingly, the diarization branch operates on the DeepFilterNet-enhanced audio, whereas the transcription branch performs only loudness normalization following the ITU-R BS.1770 loudness normalisation recommendation~\cite{ITU1770} before WhisperX inference. Loudness normalization provides a consistent recording level without aggressively modifying the spectral characteristics of the speech signal, thereby preserving the information required for robust automatic speech recognition. Alternative speech enhancement approaches were also investigated during preliminary experimentation (see Appendix~\ref{appendix:ablation}); however, based on the observed trade-off between speaker diarization and transcription performance, DeepFilterNet and loudness normalization were selected as the final preprocessing strategies for the diarization and transcription branches, respectively.

\noindent Finally, all recordings were resampled to a single-channel 16~kHz waveform to ensure compatibility with the Pyannote voice activity detection and WhisperX transcription pipelines.

\section{Architecture}
\begin{figure}[h]
    \centering
    \includegraphics[width=0.5\linewidth]{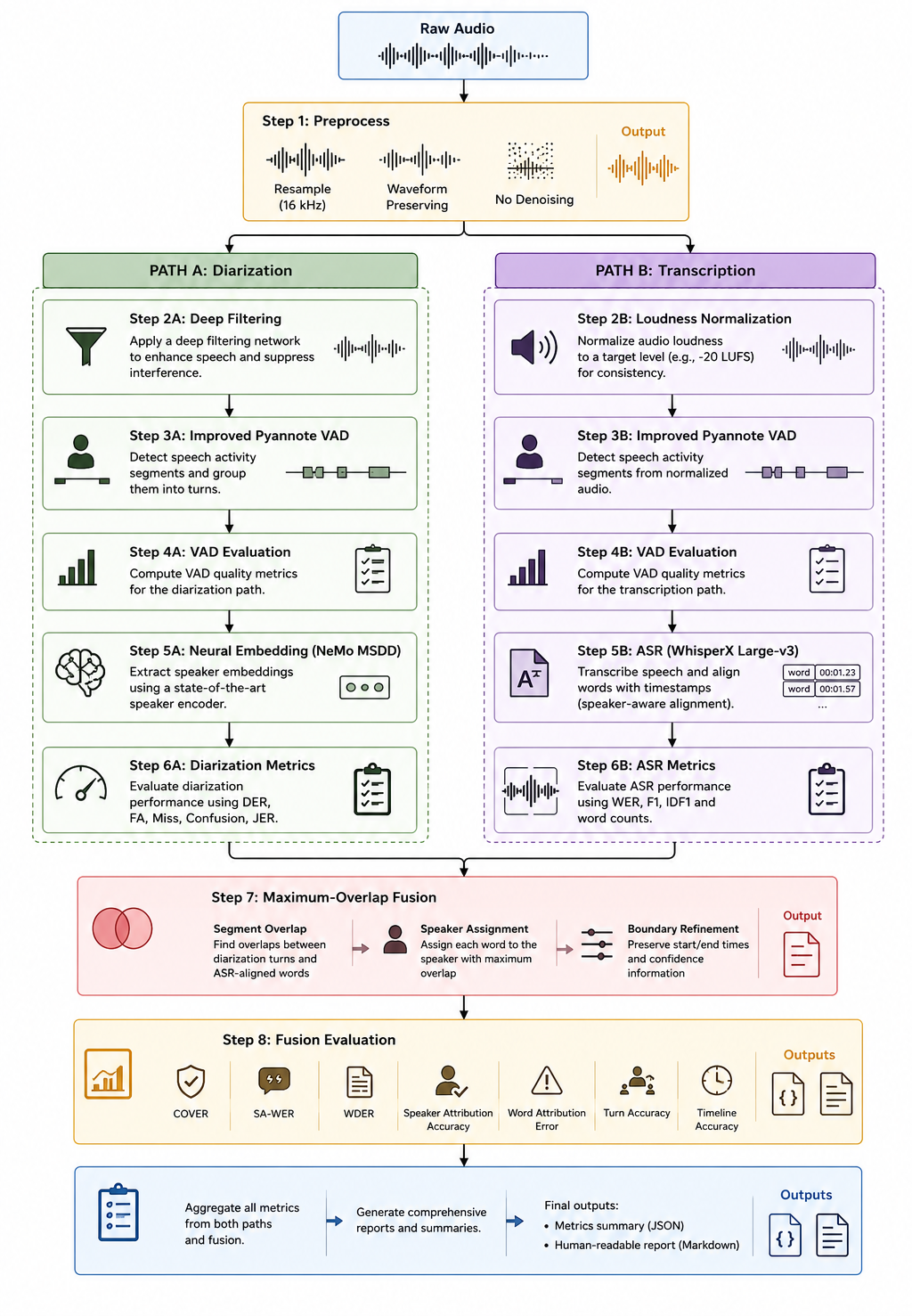}
    \caption{Overall architecture of the proposed dual-path framework for processing body-worn camera (BWC) recordings}
    \label{fig:architecture}
\end{figure}

\noindent Starting from a common preprocessed audio signal, the framework separates the processing pipeline into two independent branches, each optimized for a different downstream objective. Rather than forcing a single audio representation to satisfy both automatic speech recognition (ASR) and speaker diarization, the proposed architecture applies task-specific acoustic conditioning before recombining the outputs through a dedicated speaker-attribution module.
This design is motivated by the \textbf{Enhancement Trap} discussed in
Section~3.2. Representative acoustic conditioning experiments supporting this
design choice are presented in Appendix~\ref{appendix:ablation}.

\noindent The first branch (Path A) is optimized for speaker diarization. The preprocessed waveform is enhanced using DeepFilterNet~\cite{DeepFilterNet} to suppress environmental interference before passing through the proposed probability-guided Voice Activity Detection (VAD) pipeline. The resulting speech boundaries are subsequently supplied as externally generated speech boundaries to NVIDIA's Multi-Scale Speaker Diarization Decoder
(MSDD)~\cite{MSDD}, which employs pretrained TitaNet speaker
embeddings~\cite{TitaNet} to generate timestamped speaker segments.

\noindent The second branch (Path B) is optimized for transcription. Instead of neural enhancement, the preprocessed waveform undergoes loudness normalization following the ITU-R BS.1770 recommendation before being processed by the same improved VAD pipeline. The detected speech segments are independently transcribed using WhisperX (Large-v3)~\cite{WhisperX}, which additionally performs forced alignment to produce word-level timestamps and confidence scores.

\noindent Both branches operate on a common temporal grounding derived from the proposed VAD pipeline while preserving task-specific acoustic representations. This decoupling allows the diarization branch to benefit from aggressive noise suppression without introducing spectral artifacts into the transcription branch, thereby avoiding the conflicting optimization objectives commonly encountered when a single audio representation is shared between ASR and diarization.
This modular design also allows improvements to either branch to be incorporated
independently without modifying the remainder of the framework.

\noindent After both branches complete, the framework produces two complementary outputs: (i) a sequence of timestamped speaker diarization segments and (ii) a temporally aligned word-level transcription. These outputs are integrated through a deterministic speaker-attribution (orchestration) module, following the standard integration strategy adopted in modular ASR--speaker diarization pipelines~\cite{DiarizationLM}. For each transcribed word, the temporal overlap between the word interval and every diarization segment is computed, and the word is assigned to the speaker exhibiting the largest temporal overlap. If no temporal overlap exists, the nearest diarization segment within a predefined temporal tolerance is selected as a fallback. The resulting output is a speaker-attributed conversational transcript that preserves both the speaker identities estimated by the diarization branch and the word-level timestamps generated by the ASR branch.

\noindent The fusion stage produces both a machine-readable JSON representation containing word-level speaker attribution and a human-readable speaker-segmented conversational transcript. Finally, the framework performs stage-wise evaluation of voice activity detection, speaker diarization, automatic speech recognition, and speaker attribution, followed by aggregate reporting. This modular evaluation strategy enables independent analysis of each processing stage as well as the complete end-to-end system.

\subsection{Acoustic Conditioning and Enhancement}

The proposed dual-path architecture applies task-specific acoustic
conditioning to the speaker diarization and automatic speech recognition (ASR)
branches. Although both branches originate from the same repaired audio signal, they are optimized independently according to the requirements of their downstream tasks.

\noindent Although the acoustic conditioning differs between the two branches, both subsequently employ the proposed probability-guided Voice Activity Detection (VAD) pipeline to obtain a common temporal grounding for downstream speaker diarization and transcription.

\noindent The speaker diarization branch (Path A) processes the audio using DeepFilterNet prior to Voice Activity Detection (VAD) and speaker diarization. DeepFilterNet~\cite{DeepFilterNet} performs neural speech enhancement by suppressing stationary and non-stationary background noise while preserving speech intelligibility. The enhanced waveform therefore exhibits improved speech-to-noise characteristics, providing cleaner inputs for the subsequent VAD and speaker diarization stages to operate on acoustically cleaner speech regions. This enhancement is particularly beneficial under the challenging acoustic conditions encountered in body-worn camera recordings, where environmental noise can substantially degrade speaker diarization performance.

\noindent In contrast, the transcription branch (Path B) preserves the original spectral characteristics of the speech signal. Rather than applying neural enhancement, the repaired audio is subjected only to loudness normalization using the ITU-R BS.1770 loudness recommendation~\cite{ITU1770} with a target integrated loudness of $-20$ LUFS. Formally, the normalized waveform is obtained as

\[
x_{\mathrm{norm}}(t)=g_{\mathrm{LUFS}}\,x(t)
\]

where \(x(t)\) denotes the preprocessed waveform and \(g_{\mathrm{LUFS}}\) is the gain required to achieve the target integrated loudness. This normalization equalizes recording levels through a linear gain operation
without modifying the spectral shape of the speech signal or introducing
nonlinear enhancement artifacts.

\noindent This asymmetric conditioning strategy directly addresses the Enhancement Trap introduced in Section~3.2. Representative experiments supporting this design choice are presented in Appendix~\ref{appendix:ablation}.

\noindent The resulting dual-path conditioning strategy allows each branch to operate on an acoustic representation that is specifically optimized for its downstream objective. Consequently, speaker diarization benefits from enhanced speech, whereas transcription preserves the original spectral characteristics required for accurate lexical recognition. This complementary conditioning strategy provides the acoustic foundation for the subsequent voice activity detection, speaker diarization, automatic speech recognition, and speaker-attribution stages.

\subsection{Voice Activity Detection}

\begin{figure}[h]
    \centering
    \includegraphics[width=0.45\linewidth]{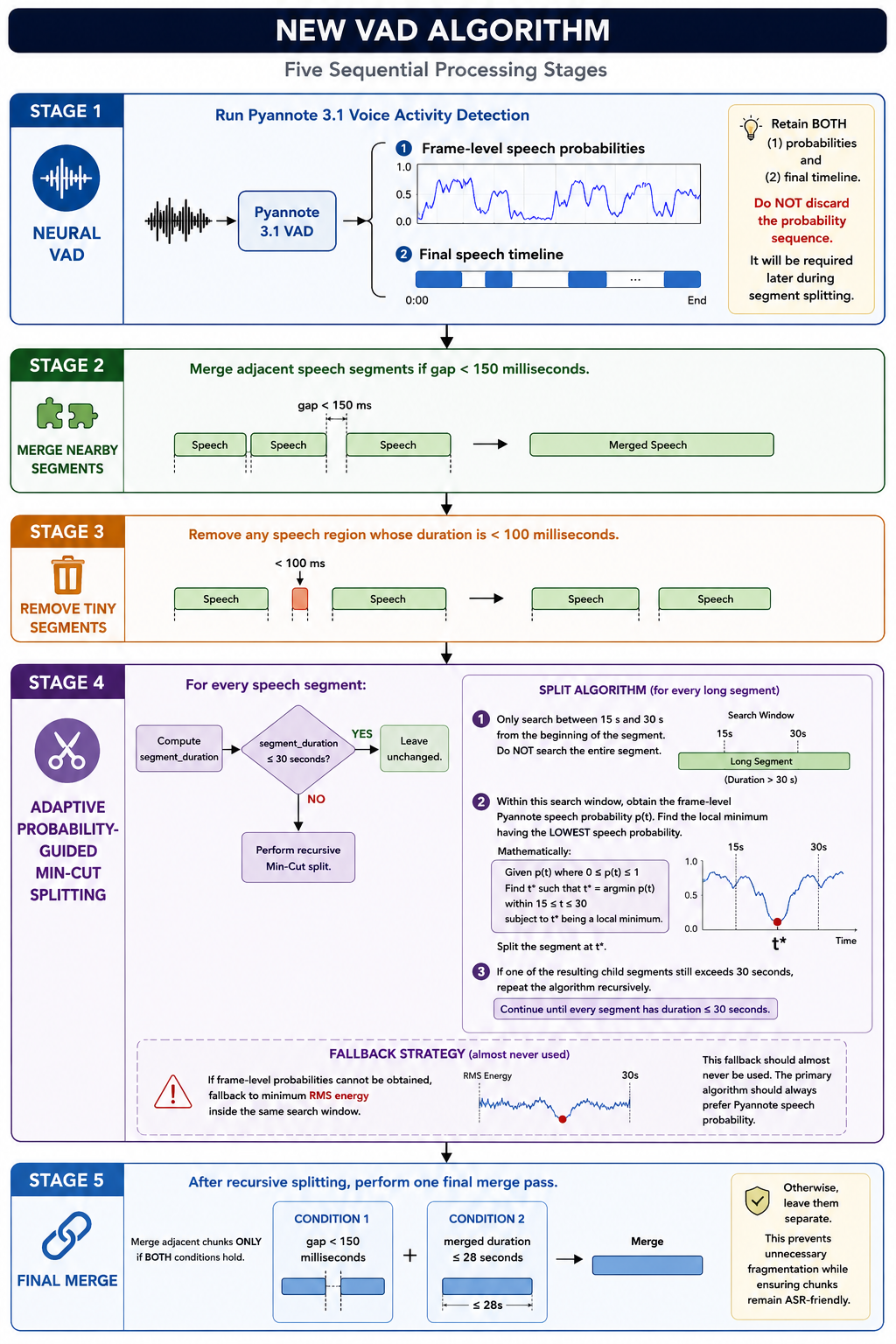}
    \caption{Overview of the proposed probability-guided Voice Activity Detection (VAD) pipeline.}
    \label{fig:vad_pipeline}
\end{figure}

\noindent The proposed Voice Activity Detection (VAD) module forms the front-end for both the diarization (Path A) and transcription (Path B) branches. We employ the Pyannote neural Voice Activity Detector~\cite{Pyannote} to estimate frame-level speech probabilities over the input audio. Unlike standard Pyannote inference, where downstream components typically operate only on the binarized speech timeline, the proposed framework explicitly preserves both the continuous speech probability sequence and the corresponding binarized speech segments, enabling probability-guided post-processing during long-segment splitting.

\noindent Let

\[
p(t)\in[0,1]
\]

denote the frame-level probability of speech predicted by Pyannote at time \(t\). The neural predictions are internally binarized using Pyannote's onset threshold, offset threshold, and hysteresis mechanism~\cite{Pyannote} to generate an initial speech timeline

\[
G=\{G_i\}_{i=1}^{N},
\qquad
G_i=[t_i^{start},t_i^{end}],
\]

where each interval corresponds to a detected speech segment.

\noindent The proposed post-processing consists of five sequential stages, illustrated in Fig.~\ref{fig:vad_pipeline}.
This probability-guided post-processing constitutes the principal methodological contribution of the proposed framework.

\paragraph{Stage 1 (Neural VAD).}
Pyannote  performs frame-level speech activity estimation and produces both the speech probability sequence \(p(t)\) and the initial binary speech timeline. Unlike the standard inference pipeline, our framework explicitly retains the frame-level probability sequence for subsequent probability-guided segmentation.

\paragraph{Stage 2 (Segment Merging).}
Adjacent speech segments are merged whenever the intervening silence duration is less than 150\,ms. This prevents unnecessary fragmentation caused by brief pauses within continuous utterances while preserving genuine speaker pauses.

\paragraph{Stage 3 (Short Segment Removal).}
Speech regions shorter than 100\,ms are discarded, as these segments are more likely to correspond to impulsive noise, transient artifacts, or isolated false-positive detections rather than meaningful speech.

\paragraph{Stage 4 (Adaptive Probability-Guided Min-Cut Splitting).}
Long speech segments are recursively divided to satisfy the maximum input duration supported by WhisperX. For every segment whose duration exceeds 30\,s, a search window between 15\,s and 30\,s from the segment onset is examined. Within this interval, the algorithm searches for the local minimum of the retained speech probability sequence. Let

\[
\mathcal{L}
=
\left\{
t \;\middle|\;
t_i^{\mathrm{start}} + 15
\le t \le
\min\!\left(t_i^{\mathrm{start}} + 30,\; t_i^{\mathrm{end}}\right),
\;
t \text{ is a local minimum of } p(t)
\right\},
\]

denote the set of all local minima within the search window. The optimal split point is then selected as

\[
t^{*}
=
\arg\min_{t\in\mathcal{L}}
p(t).
\]

\noindent The speech segment is divided at \(t^{*}\), thereby selecting the lowest-confidence speech region as the splitting point. If either resulting child segment still exceeds the maximum duration, the procedure is recursively applied until every segment satisfies the duration constraint.

\noindent Unlike the energy-based minimum-cut strategy adopted by \cite{WhisperX}, our approach directly utilizes the neural speech confidence estimated by Pyannote. Consequently, segmentation decisions are guided by the VAD model's confidence
rather than raw acoustic energy, which is expected to reduce sensitivity to
environmental acoustic interference commonly encountered in body-worn camera
recordings. Only when frame-level speech probabilities are unavailable does the algorithm fall back to the conventional minimum-RMS-energy criterion.

\paragraph{Stage 5 (Final Merge).}
Following recursive splitting, a final merge pass is performed to eliminate unnecessary fragmentation. Two adjacent segments are merged only if (i) the silence gap between them is less than 150\,ms and (ii) the merged segment remains shorter than 28\,s. Otherwise, the segments remain separate. This final constraint preserves conversational continuity while ensuring compatibility with the subsequent WhisperX transcription stage.

\noindent The resulting speech segments serve as the common temporal grounding for both processing branches. For the diarization branch, the detected speech regions are converted into Oracle\footnote{Throughout this paper, the term \emph{Oracle} refers to NeMo MSDD's \texttt{oracle\_vad=True} inference mode, in which the internal MarbleNet-based VAD is bypassed and externally supplied speech boundaries are used instead. In the proposed framework, these boundaries are generated by the probability-guided Pyannote VAD rather than by ground-truth annotations.} RTTM constraints~\cite{MSDD} for NeMo MSDD. For the transcription branch, the corresponding temporal boundaries are applied to the loudness-normalized audio prior to WhisperX transcription and forced alignment. Consequently, both downstream branches operate on a consistent temporal segmentation while preserving task-specific acoustic representations, thereby simplifying the subsequent speaker-attribution process.

\subsection{Speaker Diarization}

The speaker diarization branch operates on the speech-enhanced audio together with the master grounding segments generated by the proposed Voice Activity Detection (VAD) pipeline. Rather than using NeMo's internal MarbleNet-based Voice Activity Detector~\cite{MSDD}, the speech boundaries obtained from the proposed Pyannote-based VAD~\cite{Pyannote} are converted into an Oracle RTTM file and supplied directly to NVIDIA's Multi-Scale Speaker Diarization Decoder (MSDD)~\cite{MSDD}. Consequently, MSDD performs speaker embedding extraction and speaker diarization using the externally supplied speech regions rather than its internal VAD estimates, ensuring that all downstream stages operate on a common temporal grounding.

\noindent Let the master grounding generated by the VAD be represented as

\[
G=\{G_i\}_{i=1}^{N},
\qquad
G_i=[t_i^{start},t_i^{end}],
\]

where each interval corresponds to a detected speech segment. These intervals are converted into an Oracle RTTM file and used as the diarization constraints for MSDD, replacing the internal VAD stage while preserving the remaining diarization pipeline.

\noindent Prior to speaker embedding extraction, a binary speech mask derived from the master grounding is applied to the enhanced waveform. Unlike the default NeMo inference pipeline, speaker embedding extraction is therefore constrained exclusively to speech regions identified by the proposed VAD pipeline, preventing speaker embeddings from being extracted in regions identified as non-speech by the proposed VAD pipeline. Formally,

\[
x_{\mathrm{masked}}(t)=
x(t)\cdot m(t),
\]

where \(x(t)\) denotes the enhanced audio signal and

\[
m(t)=
\begin{cases}
1,& t\in\displaystyle\bigcup_{i=1}^{N}G_i,\\
0,&\text{otherwise}.
\end{cases}
\]

is the binary speech mask constructed from the VAD segments. This suppresses non-speech regions so that the speaker embedding extractor observes only active speech, thereby reducing the influence of background acoustic interference during downstream speaker diarization.

\noindent Speaker representation is obtained using NVIDIA's TitaNet-Large speaker encoder~\cite{TitaNet}, which generates fixed-dimensional speaker embeddings for each speech segment. These embeddings are subsequently processed by the NeMo Multi-Scale Speaker Diarization Decoder (MSDD)~\cite{MSDD}, which jointly exploits multiple temporal resolutions to refine speaker segmentation and speaker assignments. The multi-scale formulation enables robust handling of rapid speaker transitions while maintaining consistent speaker identities throughout extended conversations.
Representative experiments motivating the selection of the MSDD-based
diarization pipeline are presented in Appendix~\ref{appendix:diarization}.

\noindent The diarization output is represented as a sequence of timestamped speaker segments,

\[
D=\{(t_i^{start},t_i^{end},s_i)\}_{i=1}^{M},
\]

where \(s_i\) denotes the assigned speaker label. 

\noindent Adjacent diarization segments assigned to the same speaker are merged whenever the temporal gap between consecutive segments does not exceed 2\,s, thereby reducing unnecessary fragmentation and producing coherent conversational turns for the subsequent word-level speaker attribution stage.

\noindent The final diarization output is exported in the standard RTTM format and subsequently combined with the word-level timestamps produced by WhisperX during the speaker-attribution fusion stage.

\subsection{Speech-to-Text}

The ASR branch transcribes the LUFS-normalized audio generated by Path B using the temporally grounded speech segments generated by the proposed VAD pipeline. We employ the WhisperX framework~\cite{WhisperX}, together with OpenAI's Whisper Large-v3 model~\cite{Whisper}, to perform automatic speech recognition. Each detected speech segment is transcribed independently with \texttt{condition\_on\_previous\_text=False}, thereby preventing the transcription of one segment from influencing subsequent segments.

\noindent During preliminary experimentation, enabling contextual conditioning occasionally propagated transcription errors from acoustically challenging segments into subsequent speech segments. Since body-worn camera recordings frequently contain abrupt environmental disturbances and rapidly changing conversational contexts, disabling previous-text conditioning produced more stable segment-level transcriptions.

\noindent Let the improved VAD produce a sequence of speech segments

\[
S=\{S_i\}_{i=1}^{N},
\qquad
S_i=[t_i^{start},t_i^{end}],
\]

where each segment satisfies the maximum duration constraint imposed by the VAD post-processing. For every speech segment, WhisperX produces

\[
T_i=f_{\mathrm{ASR}}(S_i)
=
\left\{
\left(
w_j,
t_j^{start},
t_j^{end},
c_j
\right)
\right\},
\]

where \(w_j\) denotes the recognized word, \(t_j^{start}\) and \(t_j^{end}\) are the aligned word-level timestamps, and \(c_j\) represents the confidence score associated with the aligned word.

\noindent Following transcription, WhisperX performs forced alignment using language-specific phoneme alignment models~\cite{WhisperX} to refine the word-level timestamps. Forced alignment compensates for small segmentation offsets introduced during VAD and produces precise start and end times for every recognized word. For example, the word \emph{``vehicle''} may be aligned between 8.10\,s and 8.55\,s with an associated confidence score of 0.91.

\noindent The improved VAD post-processing ensures that every speech segment remains shorter than 30\,s before transcription, thereby ensuring compatibility with the Whisper inference window for long-form
recordings. 

\noindent WhisperX combines VAD-guided segmentation with phoneme-level forced alignment and has demonstrated highly competitive performance for long-form automatic speech recognition and word-level timestamp estimation~\cite{WhisperX}. Within the proposed framework, WhisperX provides both the lexical transcription and the temporally precise word-level timestamps required by the subsequent speaker-attribution fusion stage. The confidence associated with every aligned word is preserved throughout the pipeline and included in the final speaker-attributed conversational transcript.

\subsection{Speaker Attribution Fusion}

\noindent \noindent The final stage of the proposed framework integrates the outputs of the speaker diarization and automatic speech recognition branches to generate a speaker-attributed conversational transcript. Following the temporal orchestration strategy commonly employed to combine automatic speech recognition and speaker diarization outputs~\cite{DiarizationLM}, each transcribed word is assigned to the speaker whose diarization segment exhibits the greatest temporal overlap with the aligned word interval. If a word does not overlap any diarization segment, the temporally nearest speaker segment is selected as a fallback. This strategy preserves the precise word-level timestamps produced during forced alignment while combining the complementary strengths of the diarization and transcription branches.

\noindent Let the diarization output be represented as

\[
D=\left\{(t_i^{\mathrm{start}},t_i^{\mathrm{end}},s_i)\right\}_{i=1}^{M},
\]

where \(s_i\) denotes the speaker label associated with the diarization interval \([t_i^{\mathrm{start}},t_i^{\mathrm{end}}]\). Similarly, let the aligned transcription be represented as

\[
W=\left\{(w_j,t_j^{\mathrm{start}},t_j^{\mathrm{end}},c_j)\right\}_{j=1}^{N},
\]

where \(w_j\) denotes the recognized word, \(t_j^{\mathrm{start}}\) and \(t_j^{\mathrm{end}}\) are the aligned word boundaries, and \(c_j\) is the corresponding confidence score.

\noindent For every transcribed word, the temporal overlap between the word interval and every diarization segment is computed. The assigned speaker label is obtained as

\[
\hat{s}(w_j)=s_k,
\qquad
k=
\arg\max_i
\left|
[t_j^{\mathrm{start}},t_j^{\mathrm{end}}]
\cap
[t_i^{\mathrm{start}},t_i^{\mathrm{end}}]
\right|,
\]

where \(|\cdot|\) denotes the duration of the overlapping interval. Consequently, every word is assigned to the speaker occupying the largest portion of its temporal duration. Since speaker attribution depends solely on temporal overlap, the proposed fusion algorithm is deterministic and does not require any additional learned parameters or post-processing models.

\noindent If a word does not overlap any diarization segment, the speaker label of the temporally nearest diarization segment is assigned as a fallback. If no suitable diarization segment exists, the word is assigned the label \texttt{UNK}. This strategy prevents isolated alignment inconsistencies from leaving words without speaker assignments while maintaining deterministic behaviour.

\noindent After speaker assignment, adjacent words are grouped into conversational turns, thereby reducing unnecessary speaker fragmentation. Consecutive words are merged whenever they belong to the same speaker and the temporal gap between adjacent words does not exceed \(2\,\mathrm{s}\). Formally, two consecutive words \(w_i\) and \(w_{i+1}\) are merged if

\[
\hat{s}(w_i)=\hat{s}(w_{i+1})
\]

and

\[
t_{i+1}^{\mathrm{start}}-t_i^{\mathrm{end}}\le2\,\mathrm{s}.
\]

Otherwise, a new speaker turn is created. This procedure produces coherent conversational segments while preserving the original word-level timestamps.

\noindent The fusion stage generates two complementary outputs. The first is a machine-readable JSON representation containing every recognized word together with its aligned timestamps, confidence score, and assigned speaker label. The second is a human-readable conversational transcript in which adjacent words belonging to the same speaker are grouped into conversational turns. These complementary outputs provide both structured data for downstream conversational intelligence applications and an interpretable transcript suitable for manual review and subsequent language-model-based processing.

\noindent The quality of the generated speaker-attributed transcript is evaluated using the speaker attribution metrics described in Section~5. Speaker Accuracy measures the proportion of words correctly assigned to their corresponding speakers after optimal one-to-one speaker mapping. Word Attribution Error (WAE) is defined as the complement of Speaker Accuracy, while Timeline Accuracy measures the frame-level agreement between the predicted and reference speaker timelines. Together, these metrics quantify the quality of the final speaker-attributed transcript while complementing the independent ASR and speaker diarization evaluations.

\section{Results}

\subsection{Experimental Evaluation}

We evaluate the proposed dual-path conversational intelligence framework on an internally curated body-worn camera (BWC)-like dataset comprising publicly available recordings with manually annotated speaker-aware transcripts and diarization references. Each stage of the pipeline is evaluated independently using task-specific metrics before assessing the quality of the final speaker-attributed conversational transcript. 

\noindent This stage-wise evaluation enables the contribution of each processing component to be analysed separately while also assessing the overall end-to-end performance. The reported metrics are obtained directly from the evaluation scripts described in Section 4 and are summarized in Tables~\ref{tab:vad}, \ref{tab:diar}, \ref{tab:asr}, and \ref{tab:fusion}. The numerical values reported below correspond to the proposed framework.

\subsubsection{VAD}

\begin{table}[h]
\centering
\caption{Voice Activity Detection Performance}
\label{tab:vad}
\begin{tabular}{lcc}
\hline
Metric & Path A & Path B\\
\hline
Precision & 0.8448 & 0.8417\\
Recall & 0.7721 & 0.7670\\
F1-score & 0.8068 & 0.8026\\
False Alarm & 0.2991 & 0.3042\\
Miss & 0.2279 & 0.2330\\
\hline
\end{tabular}
\end{table}

\noindent Table~\ref{tab:vad} presents the VAD performance for both processing branches. The diarization branch (Path A) achieves an F1-score of 0.8068, while the transcription branch (Path B) achieves an F1-score of 0.8026. The nearly identical performance demonstrates that the proposed probability-guided post-processing remains stable under both acoustic conditioning strategies. Path A exhibits marginally higher precision and recall, suggesting that the enhanced audio provides slightly more robust speech
detection under the evaluated recording conditions, without introducing excessive false alarms. Overall, the proposed VAD provides reliable speech segmentation for both downstream tasks while enforcing the maximum-segment-length constraint required by WhisperX.

\noindent The small performance difference between the two branches further indicates that the proposed probability-guided post-processing remains largely independent of the acoustic conditioning applied before VAD, allowing a common temporal grounding to be shared across both processing paths. Representative ablation experiments are presented in Appendix~\ref{appendix:ablation}.

\subsubsection{Diarization}

\begin{table}[h]
\centering
\caption{Speaker Diarization Performance}
\label{tab:diar}
\begin{tabular}{lc}
\hline
Metric & Value\\
\hline
DER & 0.4471\\
Miss & 0.2300\\
False Alarm & 0.1399\\
Confusion & 0.0771\\
JER & 0.8424\\
\hline
\end{tabular}
\end{table}

\noindent The proposed system achieves a Diarization Error Rate (DER) of 44.71\%, comprising 23.00\% Miss, 13.99\% False Alarm, and 7.71\% Speaker Confusion. The Miss and False Alarm components constitute the largest proportion of the overall DER, indicating that missed speech regions and false speech detections contribute more significantly to the total diarization error than incorrect speaker assignments. This suggests that, under the challenging acoustic conditions of body-worn camera recordings, accurately identifying speech activity remains a greater source of error than speaker discrimination once speech has been detected.

The Diarization Error Rate (DER) is computed as

\[
\mathrm{DER}
=
\mathrm{Miss}
+
\mathrm{FalseAlarm}
+
\mathrm{Confusion},
\]

where Miss denotes reference speech not detected, False Alarm corresponds to non-speech incorrectly classified as speech, and Confusion measures speech assigned to the wrong speaker after optimal speaker mapping.

\noindent The corresponding Jaccard Error Rate (JER) is 0.8424. Unlike DER, which measures frame-level diarization errors, JER evaluates the temporal overlap between the predicted and reference speaker regions after optimal speaker matching. The relatively high JER indicates limited overlap between the predicted and reference speaker segments, reflecting the difficulty of achieving temporally accurate speaker segmentation in the presence of environmental noise, rapidly changing conversational turns, and overlapping speech.
The selection of the final diarization pipeline is further justified in
Appendix~\ref{appendix:diarization}.

\subsubsection{ASR}

\begin{table}[h]
\centering
\caption{Automatic Speech Recognition Performance}
\label{tab:asr}
\begin{tabular}{lc}
\hline
Metric & Value\\
\hline
WER & 0.3244\\
Substitutions & 84\\
Insertions & 27\\
Deletions & 46\\
Reference Words & 484\\
Hypothesis Words & 465\\
\hline
\end{tabular}
\end{table}

Word Error Rate (WER) is computed as

\[
\mathrm{WER}
=
\frac{S+D+I}{N},
\]

where \(S\), \(D\), and \(I\) denote the numbers of substitutions, deletions, and insertions, respectively, and \(N\) is the total number of reference words.

\noindent Table~\ref{tab:asr} reports the transcription performance of WhisperX using the loudness-normalized audio from Path~B. The proposed system achieves a Word Error Rate (WER) of 32.44\%, comprising 84 substitutions, 27 insertions, and 46 deletions over a reference transcript containing 484 words. The observed transcription errors primarily reflect the challenging acoustic characteristics of body-worn camera recordings, including environmental noise, overlapping speech, speaker movement, and conversational interruptions. Despite these adverse conditions, WhisperX successfully produces word-level transcriptions with temporally precise alignments, providing the timestamped lexical information required for the subsequent speaker-attribution fusion stage. During preliminary experimentation, applying neural speech enhancement to the transcription branch consistently increased transcription errors, whereas loudness normalization preserved the spectral characteristics required by WhisperX for robust lexical recognition. The final pipeline therefore performs transcription exclusively on the loudness-normalized audio while reserving speech enhancement for the diarization branch. The acoustic conditioning experiments motivating this design are discussed in
Appendix~\ref{appendix:acoustic}.

\subsubsection{Fusion}

\begin{table}[h]
\centering
\caption{Fusion Performance}
\label{tab:fusion}
\begin{tabular}{lc}
\hline
Metric & Value\\
\hline
Speaker Accuracy & 0.9027\\
Word Attribution Error (WAE) & 0.0973\\
Timeline Accuracy & 0.7053\\
\hline
\end{tabular}
\end{table}

\noindent Speaker Accuracy is computed as

\[
\mathrm{SpeakerAccuracy}
=
\frac{1}{N}
\sum_{i=1}^{N}
\mathbf{1}
\!\left(
s_i^{\mathrm{pred}}
=
s_i^{\mathrm{ref}}
\right),
\]

where \(N\) denotes the total number of transcribed words after optimal one-to-one speaker mapping, \(s_i^{\mathrm{pred}}\) and \(s_i^{\mathrm{ref}}\) denote the predicted and reference speaker labels for the \(i\)-th word, respectively, and \(\mathbf{1}(\cdot)\) is the indicator function. The proposed framework achieves a Speaker Accuracy of 90.27\%, indicating that approximately nine out of every ten transcribed words are assigned to the correct speaker.

\noindent The Word Attribution Error (WAE) is defined as the complement of Speaker Accuracy,

\[
\mathrm{WAE}
=
1-
\mathrm{SpeakerAccuracy},
\]

yielding a WAE of 9.73\%, which indicates that relatively few words are incorrectly attributed following the proposed maximum-overlap fusion process.

\noindent Timeline Accuracy evaluates the temporal consistency of the generated conversational transcript by comparing the predicted and reference active-speaker sets at every 10\,ms evaluation frame over the complete recording,

\[
\mathrm{TimelineAccuracy}
=
\frac{1}{K}
\sum_{k=1}^{K}
\mathbf{1}
\!\left(
\mathcal{A}_k^{\mathrm{pred}}
=
\mathcal{A}_k^{\mathrm{ref}}
\right),
\]

where \(K\) denotes the total number of evaluation frames, and \(\mathcal{A}_k^{\mathrm{pred}}\) and \(\mathcal{A}_k^{\mathrm{ref}}\) represent the predicted and reference sets of active speakers at frame \(k\), respectively. A frame is counted as correct only when the complete active-speaker set exactly matches the reference. The proposed framework achieves a Timeline Accuracy of 70.53\%, indicating that approximately seven out of every ten evaluation frames exhibit an exact match between the predicted and reference speaker activity. Since this metric jointly depends on the quality of speaker diarization, word-level alignment, and speaker attribution, it provides an end-to-end assessment of the temporal consistency of the generated conversational transcript.

\noindent Overall, the results indicate that the proposed maximum-overlap fusion strategy effectively integrates the complementary outputs of the diarization and transcription branches, producing a coherent speaker-attributed conversational transcript with high word-level speaker attribution accuracy while maintaining consistent temporal speaker alignment across the conversation.

\subsection{Discussion}

The experimental results validate the principal design decisions underlying the proposed dual-path conversational intelligence framework. The probability-guided VAD pipeline provided reliable speech segmentation for both processing branches while enforcing WhisperX's maximum segment-length constraint. Using a common VAD framework after branch-specific acoustic conditioning ensured that both branches operated on consistent temporal boundaries, simplifying subsequent speaker attribution.

\noindent One of the key findings of this work is the differing influence of acoustic conditioning on speaker diarization and automatic speech recognition. Our preliminary experiments (see Appendix~\ref{appendix:ablation}) demonstrated what we refer to as the \textbf{Enhancement Trap}: although neural speech enhancement improves perceptual speech quality and facilitates speaker embedding extraction, it simultaneously modifies spectral characteristics that modern ASR models rely upon for robust lexical recognition. Consequently, applying DeepFilterNet to the transcription branch increased transcription errors, whereas loudness normalization preserved the acoustic characteristics required by WhisperX while still providing consistent recording levels. These observations support the proposed dual-path architecture, in which each branch is optimized independently according to its downstream objective rather than forcing a single audio representation to satisfy both tasks.

\noindent The deterministic maximum-overlap fusion algorithm further demonstrates the effectiveness of combining complementary information from the diarization and transcription branches. The high Speaker Accuracy and low Word Attribution Error indicate that accurate word-level speaker attribution can be achieved without introducing additional trainable models beyond the diarization and ASR components. Nevertheless, the lower Timeline Accuracy illustrates that the quality of the final conversational transcript remains fundamentally dependent on the temporal precision of both speaker diarization and word alignment. Errors introduced during voice activity detection, diarization, or transcription inevitably propagate through the subsequent fusion stage, highlighting the importance of improving each upstream component.

\noindent Although the proposed framework demonstrates promising performance on challenging body-worn camera recordings, several limitations remain. The evaluation was conducted on an internally curated dataset assembled from publicly available recordings, and therefore the reported results may not directly generalize to other recording conditions or operational deployments. Furthermore, the relatively high Diarization Error Rate indicates that accurately localizing speaker boundaries under highly dynamic acoustic conditions remains a challenging problem. Environmental noise, overlapping speech, rapid speaker transitions, and short conversational turns remain challenging and provide clear directions for future research.

\section{Future Directions}

\subsection{LLM-based Speaker Attribution and Transcript Refinement}

\noindent The current framework performs speaker attribution using a deterministic maximum-overlap fusion strategy, in which each aligned word is assigned to the speaker segment exhibiting the greatest temporal overlap. While this approach is transparent, computationally efficient, and fully reproducible, it relies exclusively on temporal information and does not exploit the semantic relationships that naturally exist within multi-speaker conversations.

\noindent Recent advances have shown that Large Language Models (LLMs) can serve as effective post-processing modules for speaker diarization and speaker-attributed conversational understanding. Rather than replacing conventional speech processing components, these approaches leverage the outputs of existing automatic speech recognition and speaker diarization systems to reason about speaker identities using linguistic context, conversational structure, discourse coherence, and world knowledge.

\noindent For example, Wang \textit{et al.} proposed \textit{DiarizationLM}~\cite{DiarizationLM}, in which an LLM refines speaker assignments after diarization by jointly analysing the conversational transcript and the initial speaker labels. Their results demonstrate that semantic reasoning can resolve speaker assignment errors that are difficult to distinguish using acoustic information alone, particularly during ambiguous speaker transitions and overlapping conversational exchanges.

\noindent Similarly, Park \textit{et al.}~\cite{Park2024} introduced a contextual beam-search framework in which an LLM evaluates multiple speaker assignment hypotheses using conversational context, thereby improving speaker assignment consistency across long-form conversations.

\noindent Beyond speaker diarization, Adedeji \textit{et al.}~\cite{Adedeji2024} demonstrated that LLMs can also improve automatic speech recognition by correcting transcription errors through contextual reasoning. Although their work focuses on medical conversations, the underlying principle may also be applicable to body-worn camera recordings, where domain-specific terminology, challenging acoustic conditions, and fragmented conversational speech frequently reduce transcription quality.

\noindent These recent developments suggest a natural extension of the proposed framework. Since the current pipeline already produces a temporally aligned speaker-attributed conversational transcript, an LLM could be incorporated as a modular post-processing stage after speaker-attribution fusion. The language model could refine speaker assignments using conversational semantics while preserving the temporal structure generated by the deterministic speaker-attribution module described in Section~4.5.

\noindent Beyond correcting residual speaker-attribution errors, such semantic post-processing could enable higher-level conversational intelligence tasks, including dialogue summarization, speaker role identification (e.g., officer, civilian, witness), event extraction, question answering, and automatic incident report generation. Thus, the proposed framework provides a structured conversational representation that can serve as a foundation for future LLM-based conversational intelligence systems.

\subsection{Multi-View Speaker Representation Learning}

\noindent Although the proposed framework employs NVIDIA's TitaNet speaker embeddings~\cite{TitaNet} within the MSDD diarization pipeline~\cite{MSDD}, the quality of speaker clustering remains fundamentally dependent on the robustness of the underlying speaker representation. In challenging body-worn camera recordings, factors such as environmental noise, reverberation, overlapping speech, speaker movement, and varying recording distances can significantly degrade the discriminative power of a single speaker embedding model.

\begin{figure}[h]
    \centering
    \includegraphics[width=0.975\linewidth]{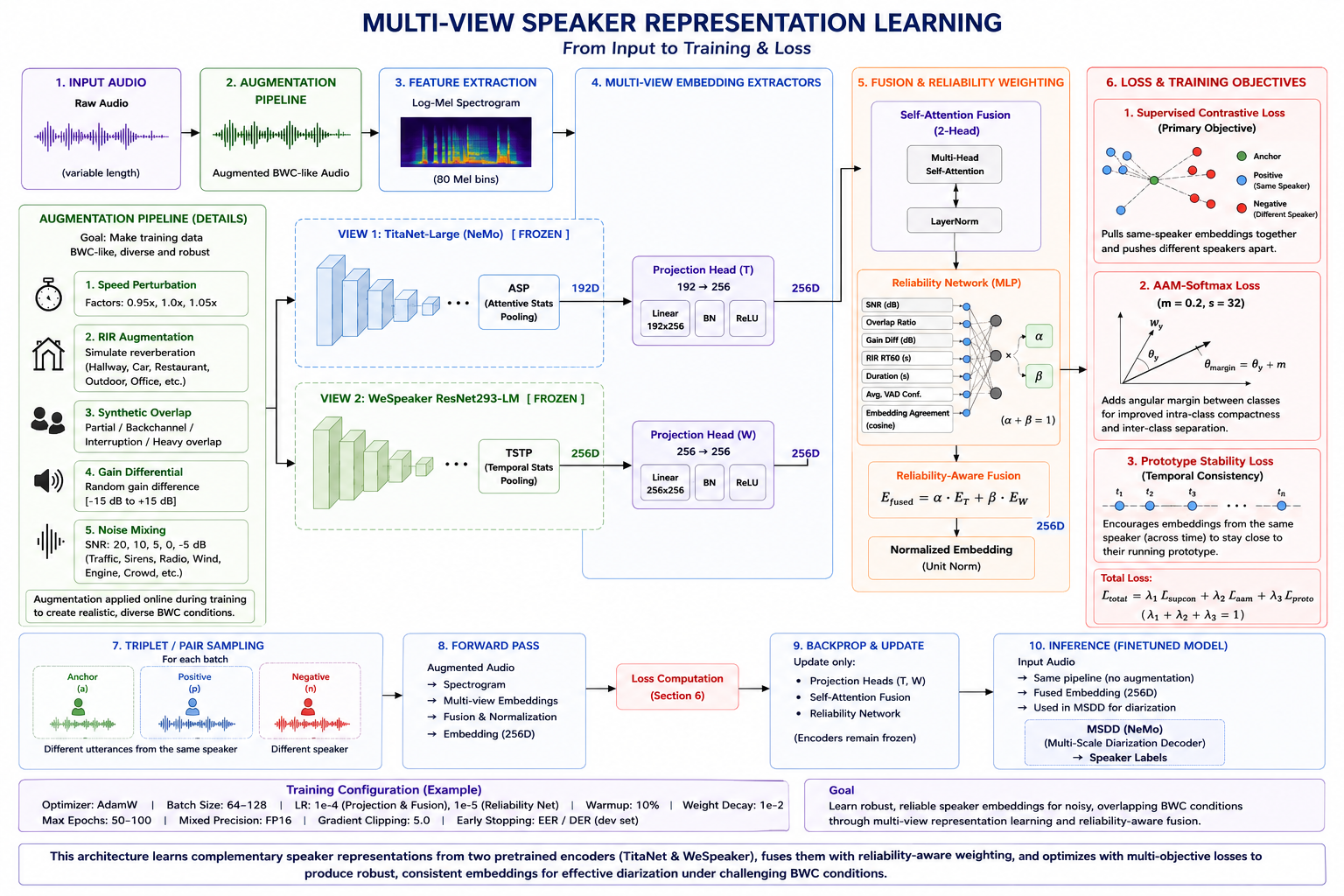}
    \label{fig:embedding_pipeline}
\end{figure}

\noindent Recent research has explored multi-view attention mechanisms for learning more robust speaker representations by capturing complementary local and global acoustic characteristics within Transformer-based speaker embedding models~\cite{Wang2022MultiViewTransformer}. Building upon this idea, a possible extension of the proposed framework is to employ multiple pretrained speaker encoders in their frozen state so that complementary acoustic representations can be extracted from the same speech segment. These embeddings could subsequently be projected into a shared representation space and fused using a lightweight self-attention mechanism to obtain a more robust speaker representation for downstream diarization.

\noindent A promising extension of the proposed framework is to replace the current single-encoder speaker representation with a reliability-aware multi-view embedding module. Multiple pretrained speaker embedding models, such as TitaNet~\cite{TitaNet}, WeSpeaker~\cite{WeSpeaker}, and related architectures, could be jointly leveraged to extract diverse speaker representations. These embeddings could then be projected into a shared latent space and adaptively fused using a lightweight attention-based module. A reliability estimation network could further assign dynamic weights based on acoustic characteristics such as signal-to-noise ratio, reverberation, overlapping speech, background noise level, and voice activity confidence. Such an adaptive fusion strategy could improve the robustness of speaker representations under challenging recording conditions while remaining compatible with the proposed dual-path framework.

\noindent Such multi-view learning strategies can also be combined with acoustic data augmentation and contrastive learning to improve the robustness of speaker embeddings under noisy recording conditions.

\noindent The resulting fused embedding could be integrated directly into the existing MSDD diarization pipeline without modifying the remainder of the proposed architecture. Since the current framework already provides robust speech segmentation through the improved VAD pipeline, the impact of improved speaker representations on downstream speaker diarization could be evaluated independently. Such a modular design preserves the advantages of the proposed dual-path framework while providing a principled direction for improving downstream speaker diarization under challenging body-worn camera recording conditions.

\subsection{Task-Specific Speaker Embedding Adaptation}

Unlike the previous subsection, which focuses on combining multiple pretrained
speaker representations, the present direction investigates adapting a single
pretrained speaker encoder to the body-worn camera domain.

\noindent The proposed framework currently employs the pretrained NVIDIA TitaNet speaker encoder~\cite{TitaNet} within the MSDD diarization pipeline~\cite{MSDD} without additional domain-specific optimization. Although TitaNet demonstrates strong generalization across diverse speech datasets, it was originally trained on a large-scale speaker corpus containing approximately 16.6 thousand speakers under recording conditions that differ substantially from body-worn camera (BWC) environments. Consequently, the learned speaker embedding space may not fully capture the acoustic variability introduced by outdoor recordings, environmental noise, radio communication, reverberation, clipping, and rapidly changing speaker positions.

\begin{figure}[h]
    \centering
    \includegraphics[width=0.975\linewidth]{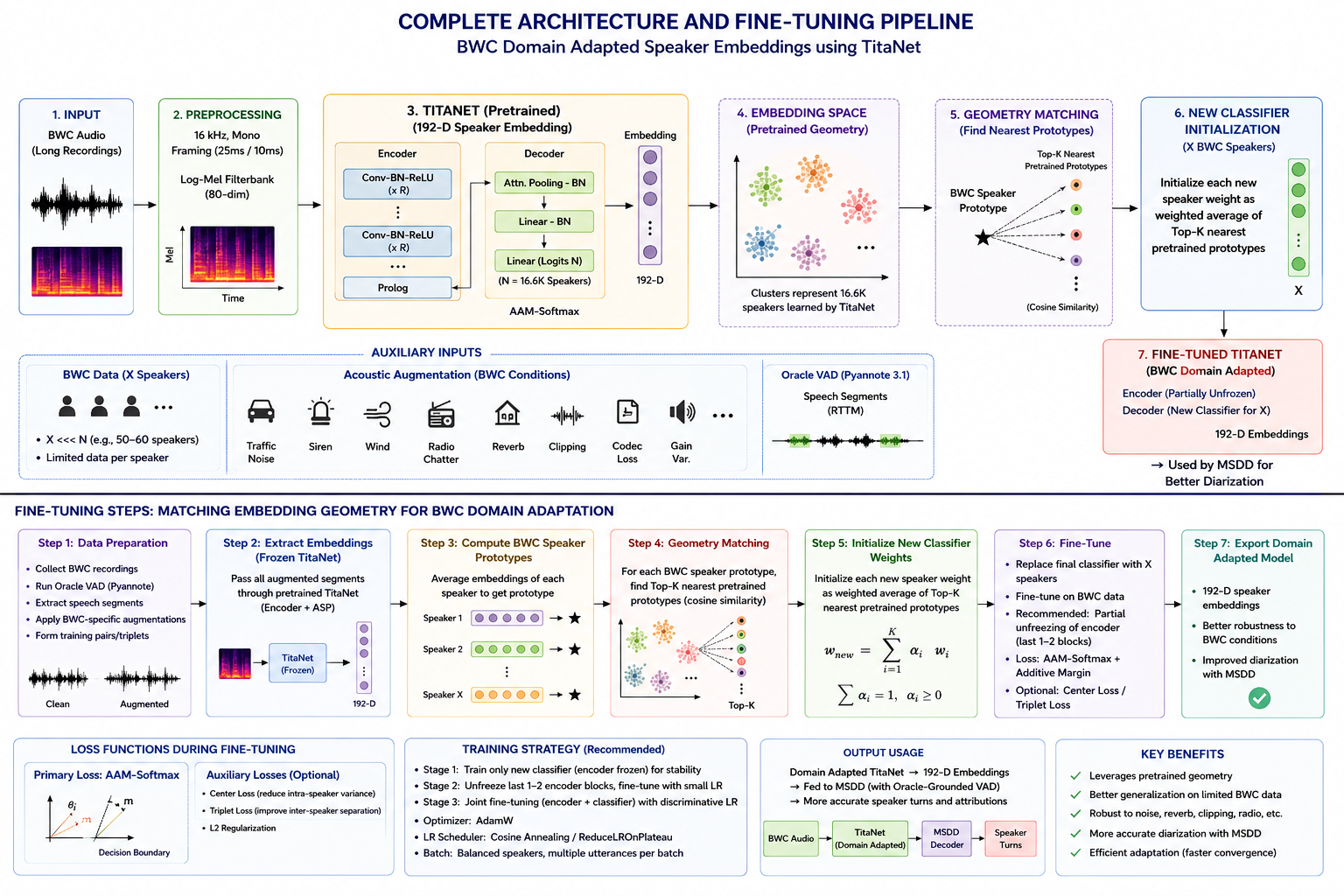}
    \label{fig:new_architecture}
\end{figure}

\noindent A promising direction for future work is to perform domain adaptation of the speaker encoder while preserving the discriminative geometry learned during large-scale pretraining. Rather than training a speaker encoder from scratch, the pretrained model can be fine-tuned using a comparatively small corpus of labeled BWC recordings. For each speaker in the adaptation dataset, multiple utterances can first be used to estimate a stable prototype embedding, providing a reliable initialization for subsequent fine-tuning while maintaining the global structure of the pretrained embedding space.

\noindent During adaptation, the original classification layer can be replaced with a task-specific classifier corresponding to the speakers present in the adaptation dataset. Training can initially be restricted to the newly initialized classifier while the pretrained encoder remains frozen. Once the classifier converges, the final layers of the encoder can be progressively unfrozen and jointly optimized using a reduced learning rate. Such staged fine-tuning enables the encoder to gradually adapt to body-worn camera acoustics while reducing catastrophic forgetting of the speaker-discriminative representations learned during large-scale pretraining.

\noindent The adaptation process can be further strengthened through domain-specific data augmentation, including traffic noise, sirens, wind, radio communication, clipping, codec degradation, reverberation, gain variation, and synthetic speaker overlap. These augmentations can be combined with metric-learning objectives, such as Additive Angular Margin Softmax (AAM-Softmax)~\cite{AAMSoftmax}, Triplet Loss~\cite{FaceNet}, or Center Loss~\cite{CenterLoss}, which are commonly used to encourage embeddings belonging to the same speaker to remain compact while simultaneously increasing the separation between different speakers under challenging acoustic conditions.

\noindent The resulting domain-adapted speaker encoder can be integrated directly into the proposed framework by replacing the pretrained TitaNet model within the MSDD diarization branch while leaving the remainder of the pipeline unchanged. Since the proposed framework already provides consistent speech segmentation through the improved VAD pipeline and task-specific acoustic conditioning through the dual-path architecture, improving the underlying speaker representation provides a principled avenue for reducing speaker confusion and enhancing downstream speaker diarization performance while preserving the modular design of the overall system.

\section{Conclusion}

This paper presented a modular dual-path conversational intelligence framework for processing body-worn camera (BWC) audio under challenging real-world acoustic conditions. Recognizing the conflicting acoustic requirements of automatic speech recognition (ASR) and speaker diarization, the proposed framework is designed to optimize each task independently through dedicated processing branches while maintaining a common temporal representation for subsequent speaker attribution. The resulting architecture combines task-specific acoustic conditioning, an improved probability-guided Voice Activity Detection (VAD) pipeline, NeMo Multi-Scale Speaker Diarization Decoder (MSDD), WhisperX transcription with forced alignment, and a deterministic maximum-overlap fusion strategy to generate a temporally aligned, speaker-attributed conversational transcript.

\noindent A key contribution of this work is the introduction of a probability-guided VAD post-processing framework that preserves frame-level speech confidence throughout segmentation and uses this information to perform adaptive recursive min-cut splitting of long speech regions. Unlike conventional energy-based segmentation strategies, the proposed approach performs splitting directly on neural speech confidence estimates while simultaneously preserving a common temporal grounding for both the diarization and transcription branches. In addition, this work identifies what we refer to as the \textbf{Enhancement Trap}, demonstrating that speech enhancement and automatic speech recognition exhibit fundamentally different acoustic requirements. These observations motivate the proposed dual-path architecture, in which each branch receives an audio representation specifically optimized for its downstream objective.

\noindent The modular design of the proposed framework enables every processing stage to generate auditable intermediate artifacts together with independent evaluation metrics. Voice activity detection, speaker diarization, automatic speech recognition, and speaker-attribution fusion are evaluated independently, providing insight into the contribution of each component while facilitating reproducibility and future experimentation. Furthermore, the modular architecture allows individual components to be replaced or improved without requiring changes to the remainder of the pipeline.

\noindent Experimental evaluation on the curated body-worn camera evaluation dataset demonstrates the effectiveness of the proposed framework. The improved VAD pipeline provides stable speech segmentation for both processing branches, while task-specific acoustic conditioning enables the diarization branch to benefit from enhanced speech and the transcription branch to preserve the acoustic characteristics required for robust lexical recognition. The proposed maximum-overlap fusion strategy further produces a speaker-attributed conversational transcript with approximately 90\% Speaker Accuracy and a Word Attribution Error below 10\%, demonstrating that complementary outputs from speaker diarization and automatic speech recognition can be effectively integrated into a coherent conversational timeline.

\noindent Although the current evaluation is limited to an internally collected body-worn camera dataset, the proposed framework provides a flexible foundation for future research in conversational intelligence. The modular pipeline can naturally accommodate future advances in Large Language Models, multi-view speaker representation learning, and domain-adaptive speaker embedding optimization while preserving the overall system architecture. These extensions offer promising directions for improving robustness, speaker attribution, and higher-level conversational understanding in real-world body-worn camera recordings.

\noindent Overall, this work demonstrates that combining task-specific acoustic conditioning, probability-guided speech segmentation, modular speaker diarization, robust transcription, and principled word-level speaker attribution provides an effective, interpretable, and reproducible framework for conversational analysis of body-worn camera audio. By addressing the conflicting acoustic requirements of speaker diarization and automatic speech recognition through a unified yet modular architecture, the proposed framework establishes a strong baseline for future research in speaker-aware conversational intelligence and practical speech analytics for complex real-world acoustic environments.

\newpage

\bibliographystyle{IEEEtran}
\bibliography{references}

\newpage

\appendix
\section{Baseline Pipeline Comparison}
\label{appendix:baseline}

To place the proposed framework in context, we compare it with the original single-path end-to-end pipeline used as the initial reference implementation. The original reference implementation consisted of a conventional single-path pipeline comprising Silero Voice Activity Detection (VAD), WhisperX Large-v3 for transcription and forced alignment, Pyannote 3.1 speaker diarization, and WhisperX's default speaker-assignment procedure. In contrast, the proposed framework introduces task-specific acoustic conditioning through the dual-path architecture, probability-guided Pyannote VAD, Oracle-grounded NeMo Multi-Scale Speaker Diarization Decoder (MSDD) using TitaNet speaker embeddings, and a deterministic maximum-overlap speaker-attribution module.

\begin{table}[H]
\centering
\caption{Comparison between the original baseline pipeline and the proposed framework.}
\label{tab:baseline}
\begin{tabular}{lcc}
\hline
Component & Baseline Pipeline & Proposed Framework\\
\hline
Audio Processing & Single-path & Dual-path (ASR / Diarization)\\
Speech Enhancement & None & DeepFilterNet (Path A only)\\
ASR Conditioning & Raw audio & $-20$ LUFS loudness normalization\\
Voice Activity Detection & Silero VAD & Pyannote VAD + probability-guided splitting\\
Speaker Diarization & Pyannote 3.1 & Oracle-grounded NeMo MSDD + TitaNet\\
Speaker Attribution & WhisperX assignment & Maximum-overlap fusion\\
Evaluation & End-to-end only & Stage-wise evaluation (VAD, ASR, Diarization, Fusion)\\
\hline
\end{tabular}
\end{table}

Table~\ref{tab:baseline} summarizes the principal architectural differences between the two systems. The proposed framework replaces the single shared audio representation with independent processing paths optimized for speaker diarization and automatic speech recognition. Furthermore, each stage of the pipeline produces intermediate artifacts together with dedicated evaluation metrics, enabling detailed analysis of the contribution of individual processing components.

\begin{table}[H]
\centering
\caption{Quantitative comparison between the original baseline pipeline and the proposed framework.}
\label{tab:baseline_results}
\begin{tabular}{lcc}
\hline
Metric & Baseline & Proposed\\
\hline
VAD Precision & \textbf{0.8684} & 0.8448\\
VAD Recall & 0.7511 & \textbf{0.7721}\\
VAD F1-score & 0.8055 & \textbf{0.8068}\\
DER & 0.4516 & \textbf{0.4471}\\
Speaker Confusion & 0.0949 & \textbf{0.0771}\\
WER & \textbf{0.2968} & 0.3244\\
Speaker Accuracy & 0.8981 & \textbf{0.9027}\\
Timeline Accuracy & 0.6646 & \textbf{0.7053}\\
\hline
\end{tabular}
\end{table}

The comparison highlights the primary design objective of the proposed framework. Although the proposed architecture exhibits a modest increase in Word Error Rate relative to the original single-path pipeline, it substantially reduces speaker confusion, improves timeline accuracy, and introduces several capabilities absent from the baseline, including probability-guided speech segmentation, Oracle-grounded speaker diarization, deterministic word-level speaker attribution, and stage-wise evaluation. These additions transform a conventional speech processing pipeline into a modular conversational intelligence framework while preserving reproducibility and facilitating future extensions.

\section{Component-wise Ablation and Design Justification}
\label{appendix:ablation}

The final framework presented in this paper was derived through a series of
controlled experiments investigating acoustic conditioning, voice activity
detection, speaker diarization, and speaker-attribution strategies for
body-worn camera (BWC) recordings. Rather than resulting from a single design
choice, the proposed architecture emerged through successive evaluation of
alternative processing strategies under realistic recording conditions.

This appendix summarizes only those experiments that directly influenced the
final architectural decisions. Implementation details and exploratory experiments that did not directly
influence the final architecture are intentionally omitted.

\subsection{Evolution of Proposed Framework}
\label{appendix:evolution}

\begin{figure}[!t]
    \centering
    \includegraphics[width=0.95\linewidth]{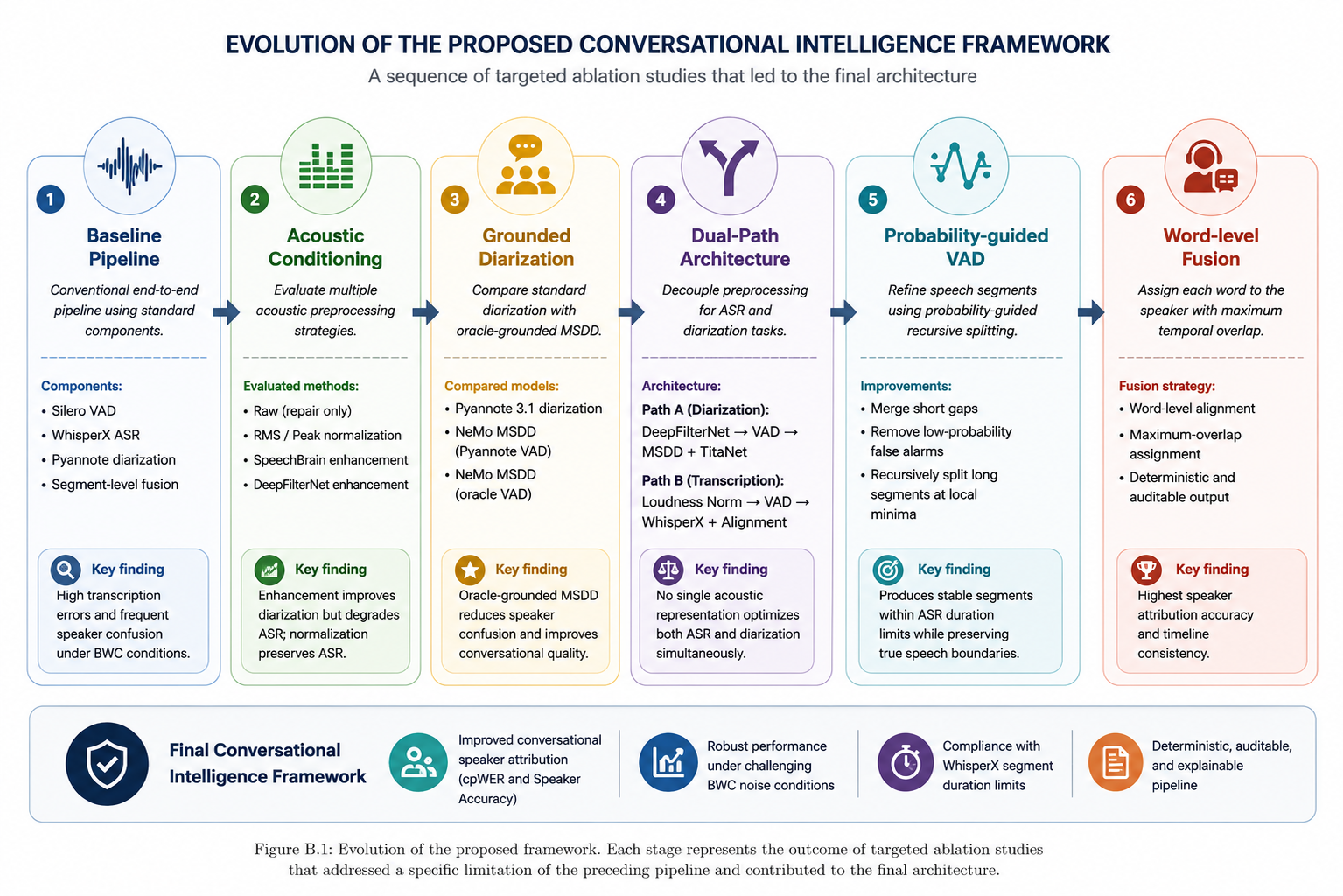}
    \caption{Evolution of the proposed framework. Only the principal architectural
    milestones that directly contributed to the final system are shown.
    Each milestone represents the outcome of a controlled experimental evaluation
    rather than an implementation iteration.}
    \label{fig:ablation_evolution}
\end{figure}

Figure~\ref{fig:ablation_evolution} summarizes the evolution of the proposed
framework. Beginning with a conventional single-path speech processing pipeline, the proposed framework progressively evolved through a series of experimentally validated architectural refinements, including task-specific acoustic conditioning, Oracle-grounded speaker diarization, probability-guided voice activity detection, and deterministic speaker-attribution fusion.

The principal experimental milestones are summarized below.

\begin{enumerate}
    \item \textbf{Acoustic conditioning.} Initial experiments evaluated several
    preprocessing strategies, including conventional normalization techniques,
    DeepFilterNet-based enhancement, and SpeechBrain enhancement, in order to
    determine their respective effects on automatic speech recognition and speaker diarization.

    \item \textbf{Speaker diarization.} The performance of conventional
    Pyannote-based diarization was subsequently compared with Oracle-grounded
    NeMo Multi-Scale Speaker Diarization Decoder (MSDD), thereby isolating the effect of externally supplied speech boundaries.

    \item \textbf{Dual-path processing.} The conflicting behaviour observed
    between speech enhancement and transcription motivated the introduction of
    separate processing branches for automatic speech recognition and speaker
    diarization.

    \item \textbf{Probability-guided speech segmentation.} The final architectural refinement
    introduced probability-guided recursive segment splitting, allowing the
    temporal segmentation produced by Pyannote to satisfy WhisperX duration
    constraints while preserving low-confidence transition regions.

    \item \textbf{Speaker attribution.} A deterministic word-level maximum-overlap fusion strategy was introduced to generate the final speaker-attributed conversational transcript, replacing earlier segment-level speaker assignment strategies explored during development.
\end{enumerate}

Collectively, these studies indicate that the final architecture did not arise
from a single architectural modification but rather from a sequence of
component-wise investigations. Each major design decision was retained only
after experimental evaluation indicated that it addressed a specific limitation
identified in the preceding configuration. The resulting framework therefore
reflects a combination of experimentally justified design choices rather than a
single end-to-end architectural redesign.

\subsection{Acoustic Conditioning Ablation}
\label{appendix:acoustic}

The first component investigated during the development of the proposed framework
was acoustic conditioning. Since both speaker diarization and automatic speech
recognition operate directly on the input waveform, the choice of preprocessing
strategy has a significant influence on downstream performance. Conventional
speech processing pipelines typically assume that improving perceptual speech
quality benefits all subsequent tasks equally. However, our preliminary
experiments demonstrated that this assumption does not hold for body-worn camera
(BWC) recordings.

To investigate this behaviour, several preprocessing strategies were evaluated,
including loudness normalization, DeepFilterNet neural speech enhancement, and
SpeechBrain enhancement. The objective was to determine whether a single acoustic representation could simultaneously provide optimal performance for both automatic speech recognition and speaker diarization.

\begin{table}[H]
\centering
\caption{Representative acoustic conditioning experiments performed during
framework development.}
\label{tab:acoustic_ablation}
\begin{tabular}{lccp{8.5cm}}
\hline
Configuration & WER & DER & Observation\\
\hline

Loudness Normalization as baseline\footnotemark[1] &
0.2975 &
0.4114 &
Provided the best balance between transcription and speaker diarization while preserving the original acoustic characteristics required for ASR.\\

DeepFilterNet (global)\footnotemark[2] &
0.3698 &
0.4408 &
Improved downstream speaker diarization but noticeably increased
transcription errors when applied throughout the complete pipeline.\\

SpeechBrain enhancement &
0.4504 &
0.5603 &
Produced the largest degradation in both transcription and diarization performance.\\

Full DeepFilterNet pipeline &
0.4215 &
0.6267 &
Global enhancement proved unsuitable for end-to-end conversational intelligence despite producing perceptually cleaner speech.\\

\hline
\end{tabular}
\end{table}

\footnotetext{The baseline presented in Appendix~\ref{appendix:baseline} corresponds to an independent comparison between existing open-source pipelines. In contrast, the experiments reported here were conducted using custom intermediate architectures developed during the design of the proposed framework.}

The experiments consistently revealed a trade-off between transcription accuracy
and speaker diarization performance, referred to throughout this work as the
\textbf{Enhancement Trap}. Although neural speech enhancement produced speech that was perceptually cleaner
for human listeners, applying the enhanced waveform throughout the complete
processing pipeline consistently increased transcription errors while altering
downstream speaker diarization behaviour. These observations indicate that
improvements in perceptual speech quality do not necessarily translate into
improvements in downstream machine perception.

For the evaluated recordings, no single acoustic representation consistently provided the best performance for both downstream tasks. While DeepFilterNet effectively suppresses environmental interference, the
resulting acoustic modifications were empirically observed to alter the acoustic characteristics presented to downstream speaker diarization and automatic speech recognition models.
Conversely, loudness normalization preserves the original spectral structure of
the speech signal while providing a consistent recording level for automatic
speech recognition.

These observations directly motivated the task-specific acoustic conditioning
strategy adopted in the final framework. DeepFilterNet is therefore applied
exclusively within the speaker diarization branch, whereas the transcription
branch operates on the loudness-normalized waveform. By allowing each branch to
operate on an acoustic representation better suited to its respective
objective,
the proposed dual-path architecture resolves the observed conflict between
speaker diarization and automatic speech recognition while preserving a common
temporal grounding through the shared VAD pipeline.
\stepcounter{footnote}
\footnotetext{The "Full DeepFilterNet pipeline" refers to an intermediate experimental configuration evaluated during framework development and should not be interpreted as the final proposed architecture.}

\subsection{Voice Activity Detection Design Study}
\label{appendix:vad}

Following the selection of task-specific acoustic conditioning, the next stage
of development focused on improving the robustness of voice activity detection
(VAD). Since both speaker diarization and automatic speech recognition depend on
accurate speech segmentation, errors introduced during VAD inevitably propagate
through the remainder of the conversational intelligence pipeline.
Consequently, several alternative VAD strategies were evaluated to determine
their suitability for long-form body-worn camera recordings.

The initial implementation employed Silero VAD owing to its lightweight design
and real-time inference capability. Although Silero produced competitive
frame-level detection accuracy, qualitative inspection revealed that long speech
segments were frequently fragmented, while abrupt environmental noise occasionally
introduced false-positive speech regions. These observations motivated the
adoption of the neural Pyannote VAD model, which provides frame-level speech
posterior probabilities rather than binary speech decisions alone.

Unlike conventional inference pipelines, the proposed framework explicitly
retains these probability estimates after speech detection. This enables the
recursive probability-guided splitting strategy introduced in Section~4.3,
allowing long speech segments to be divided at local minima of the speech
posterior rather than according to fixed-duration windows or minimum-energy
criteria.
\begin{table}[H]
\centering
\caption{Frame-level comparison between the initial Silero-based VAD and the proposed probability-guided VAD.}
\label{tab:vad_ablation}
\begin{tabular}{lcc}
\hline
Metric & Silero Baseline & Proposed VAD\\
\hline
Precision & 0.8684 & 0.8448\\
Recall & 0.7511 & 0.7721\\
F1-score & 0.8055 & 0.8068\\
\hline
\end{tabular}
\end{table}

\noindent Table~\ref{tab:vad_ablation} compares the proposed probability-guided VAD with the initial Silero-based implementation using conventional frame-level evaluation metrics. While both methods achieve comparable detection performance, the proposed VAD attains a slightly higher F1-score (0.8068 versus 0.8055). More importantly, these metrics do not evaluate the recursive probability-guided segmentation strategy introduced in the proposed framework, whose primary objective is to generate reliable temporal grounding for downstream speaker diarization and automatic speech recognition rather than to maximize frame-level VAD performance alone.

\noindent Unlike conventional VAD pipelines, the proposed framework explicitly preserves the frame-level speech posterior estimated by Pyannote and employs probability-guided recursive splitting to divide long speech segments at low-confidence regions. This enables the generated speech segments to simultaneously satisfy WhisperX duration constraints while maintaining a common temporal grounding shared by both processing branches. Consequently, the VAD module serves not only as a speech detector but also as the component responsible for establishing consistent temporal boundaries throughout the complete conversational intelligence pipeline.

\noindent The recursive probability-guided splitting strategy was introduced primarily to improve the robustness and reproducibility of long-form speech segmentation rather than to optimize conventional frame-level VAD metrics. Earlier exploratory experiments identified long uninterrupted speech regions as a practical limitation for downstream processing, particularly for WhisperX transcription under challenging recording conditions. Consequently, the proposed post-processing was designed as a deterministic preprocessing mechanism that produces reproducible temporal segmentation even when its impact is not fully reflected by conventional frame-level evaluation metrics for a particular recording.

\noindent The resulting probability-guided VAD therefore provides more than a conventional speech detector. In addition to identifying speech activity, it establishes the common temporal grounding used by both the speaker diarization and transcription branches, enabling consistent downstream processing while preserving compatibility with long-form body-worn camera recordings.

\subsection{Speaker Diarization Ablation}
\label{appendix:diarization}

The final framework employs NVIDIA's Multi-Scale Speaker Diarization Decoder
(MSDD) together with TitaNet speaker embeddings rather than the standard
Pyannote speaker diarization pipeline. This design decision emerged from a
series of controlled experiments investigating whether improvements in
conversational speaker attribution necessarily correspond to reductions in
frame-level diarization error.

Early experiments adopted the Pyannote speaker diarization pipeline owing to its
strong performance on public benchmarks and ease of deployment. Although
Pyannote consistently achieved competitive Diarization Error Rate (DER), error
analysis revealed that conversational speaker attribution remained limited by
speaker confusion, particularly during rapid conversational exchanges and
overlapping speech. These observations motivated the investigation of NeMo's
oracle-grounded MSDD architecture, which combines multi-scale TitaNet speaker
embeddings with a neural decoder capable of modelling speaker transitions across
multiple temporal resolutions.

The objective of these experiments extended beyond minimizing frame-level
diarization error. Since the ultimate goal of the proposed framework is to
produce an accurate speaker-attributed conversational transcript, the
experiments focused on identifying the architectural factors responsible for
stable speaker segmentation under challenging body-worn camera recording
conditions.

\begin{table}[H]
\centering
\caption{Representative comparison of the principal diarization architectures evaluated during framework development.}
\label{tab:diar_ablation}
\begin{tabular}{lccp{5.8cm}}
\hline
Configuration & DER & Speaker Confusion & Principal Observation\\
\hline

Pyannote baseline\footnotemark &
0.411 &
0.110 &
Competitive frame-level diarization performance, but speaker confusion during
rapid speaker transitions limited conversational consistency.\\

Oracle-grounded MSDD &
0.449 &
0.078 &
Using externally generated speech boundaries substantially reduced speaker
confusion despite a slight increase in overall DER.\\

Final framework &
0.447 &
0.077 &
Selected as the final diarization architecture owing to reduced speaker
confusion while maintaining competitive overall diarization performance.\\

\hline
\end{tabular}
\end{table}

\footnotetext{The Pyannote baseline corresponds to an independent evaluation of the diarization backend under a separate controlled experimental setting and should not be interpreted as the baseline presented in Appendix~\ref{appendix:baseline}.}

\noindent Table~\ref{tab:diar_ablation} demonstrates that the proposed diarization architecture was not selected solely on the basis of overall Diarization Error Rate (DER). Although the Oracle-grounded MSDD configuration exhibits a slightly higher DER than the Pyannote baseline, it substantially reduces speaker confusion (0.110 to 0.078), which is more critical for producing coherent speaker-attributed conversational transcripts. Consequently, the final diarization architecture was selected based on its downstream contribution to conversational intelligence rather than on frame-level diarization metrics alone.

\noindent Additional experiments further demonstrated that the quality of the voice activity detection front-end has a significant influence on downstream speaker diarization. Supplying externally generated speech boundaries through Oracle VAD consistently reduced speaker confusion by constraining speaker embedding extraction to reliable speech regions. These observations indicate that accurate temporal grounding is a key factor influencing downstream diarization performance under challenging body-worn camera recording conditions.

\noindent These observations directly motivated the final diarization architecture adopted in this work. Rather than replacing the diarization backend itself, the proposed framework focuses on improving the quality of the temporal speech grounding supplied to MSDD through the probability-guided VAD pipeline while retaining TitaNet speaker embeddings and the multi-scale speaker decoder. This modular design allows future improvements in speech segmentation or speaker representation learning to be incorporated independently without modifying the remainder of the framework.

\noindent Overall, the speaker diarization study demonstrates that the proposed framework optimizes speaker segmentation for downstream conversational intelligence rather than for minimizing DER alone. The adoption of Oracle-grounded MSDD with TitaNet embeddings therefore represents a design choice motivated by reduced speaker confusion, improved temporal consistency, and seamless integration within the proposed dual-path architecture.

\subsection{Dual-Path Architecture Ablation}
\label{appendix:dualpath}

\begin{table}[H]
\centering
\caption{Representative preprocessing configurations evaluated during the development of the proposed dual-path architecture.}
\label{tab:dualpath_design}
\begin{tabular}{p{3.5cm}p{4cm}p{5.5cm}}
\hline
Configuration & Preprocessing Strategy & Principal Observation\\
\hline

Single-path &
Identical preprocessing for ASR and diarization &
Simplifies the pipeline but forces both tasks to operate on the same acoustic representation.\\

Global enhancement &
DeepFilterNet applied before both ASR and diarization &
Reduced background interference but consistently degraded transcription
accuracy.\\

Global normalization &
Loudness normalization before both ASR and diarization &
Maintained reliable transcription but provided less favourable conditions for
speaker diarization.\\

\textbf{Dual-path architecture} &
DeepFilterNet for diarization and loudness normalization for ASR &
Allows each downstream task to operate on an acoustic representation better suited for its respective objective while maintaining common temporal grounding.\\

\hline
\end{tabular}
\end{table}

\noindent The preceding acoustic conditioning experiments demonstrated that the acoustic representation most suitable for speaker diarization differs from that required for automatic speech recognition. While neural speech enhancement generally benefited speaker diarization, it degraded transcription accuracy when applied throughout the complete pipeline. Conversely, loudness normalization preserved transcription performance but provided less favourable acoustic conditions for speaker diarization.

\noindent These findings constitute the primary empirical evidence supporting the Enhancement Trap discussed in Section~3.2 and motivate the separation of the processing pipeline into independent acoustic branches.

\noindent Several architectural configurations were therefore evaluated to determine whether separating the preprocessing strategies for speaker diarization and automatic speech recognition could resolve the conflicting behaviour observed during the acoustic conditioning experiments while maintaining a unified conversational intelligence pipeline.

\noindent These observations directly motivated the separation of the framework into two independent processing branches. The diarization branch operates exclusively on DeepFilterNet-enhanced audio to improve speaker discriminability, whereas the transcription branch preserves the original spectral characteristics through loudness normalization before WhisperX inference. Both branches subsequently share the common temporal grounding generated by the proposed probability-guided VAD pipeline, ensuring consistent downstream alignment while allowing each task to operate on an acoustically appropriate representation.

\noindent The resulting architecture is inherently modular. Because the two processing branches become independent immediately after the shared VAD stage, future improvements in speech enhancement, speaker embedding learning, speaker diarization, or automatic speech recognition can be incorporated into either branch without requiring modifications to the remainder of the framework.

\noindent Overall, the dual-path architecture emerged as the logical outcome of the preceding ablation studies rather than as an independently motivated design decision. By allowing speaker diarization and automatic speech recognition to operate on task-specific acoustic representations while preserving a common temporal grounding, the proposed framework addresses the conflicting optimization requirements identified during framework development and establishes a modular foundation for future conversational intelligence systems.

\end{document}